\documentclass[12pt]{article}
\usepackage{amsmath,amssymb,array}
\numberwithin{equation}{section}

\def \p{\partial}
\def \nn{\nonumber}
\def \l{\langle}
\def \r{\rangle}
\def \bf{\mathbf}

\begin{document}

\begin{titlepage}
\renewcommand{\thefootnote}{\fnsymbol{footnote}}

\begin{center}

\begin{flushright}
hep-th/0607188
\end{flushright}
\vspace{2cm}

\textbf{\Large{ZZ Brane Decay in $D$ Dimensions}}\\[1.5cm]

\large{Ya-Li He}$^{\,\mathrm{a}}$     \hspace{0.2cm} and
\hspace{0.2cm}
\large{Peng Zhang}$^{\,\mathrm{b}}$   \\[1cm]

$^{\mathrm{a}}$ \emph{Department of Physics, Peking University, \\
                      Beijing 100871, P. R. China} \\[0.2cm]
                \textsf{ylhe@pku.edu.cn}

\vspace{0.8cm}

$^{\mathrm{b}}$ \emph{Institute of Theoretical Physics, Chinese Academy of Science, \\
                        P.O. Box 2735, Beijing 100080, P. R. China} \\[0.2cm]
                  \textsf{pzhang@itp.ac.cn}

\end{center}
\vspace{1cm}

\centerline{\textbf{Abstract}}\vspace{0.5cm}

In this paper we consider the ZZ brane decay in a $D$-dimensional
background with a linear dilaton and a Liouville potential switched
on. We mainly calculate the closed string emission rate during the
decay process. For the case of a spacelike dilaton we find a similar
Hagedorn behavior, in the closed string UV region, with the brane
decay in the usual 26d and 2d bosonic string theory. This means that
all of the energy of the original brane converts into outgoing
closed strings. In the IR region the result is finite. We also give
some comments about the case that the dilaton is null.

\end{titlepage}
\setcounter{footnote}{0}


\section{Introduction}
String theory in time-dependent backgrounds is one of the most
important questions that need to be understood. Dealing with general
time-dependent backgrounds is still out of our abilities. Therefore
simple solvable models, capturing some important aspects of the
realistic physics, are laboratories for theoretical ideas and tools,
which may give us some valuable lessons in more general and more
difficult cases.

Among these the open string tachyon condensation, initiated by Sen
\cite{Sen1, Sen2, Sen3}, is a class of important models which are
intensely studied in the past few years. For a comprehensive
review in this subject with a complete list of references, see
\cite{Review}. These works give us many surprises and insights
into the tachyon physics. When coupling to the closed strings, a
time-dependent open string field configuration, such as the
rolling tachyon solution, on an unstable D-brane acts as a
time-dependent source of various closed string fields, and
produces closed string radiation \cite{Chen, Larsen, Rey, Lambert,
UV}. By studying various emission processes, people realize that
that there is a new kind of open/closed string duality
\cite{Sen03}. Recently the authors of \cite{Rey06} carefully study
the subtle points in the modular transformation which is related
with unitarity and channel duality. The study of open string
tachyon also leads to the ``reloading of the matrix''
\cite{McG03,KMS03}, which gives a deep holographic understanding
of the $c=1$ matrix model, identifying it as the world-line theory
of the ZZ brane low energy dynamics. This holographic viewpoint
has also been extended to the supersymmetric cases \cite{Taka03,
new-hat, McMV}.

In the calculation of the closed string radiation rate, people find
a Hagedorn divergence both in 26d \cite{Lambert} and 2d \cite{KMS03}
string theory. In \cite{UV} the authors consider the background with
a linear dilaton, and they find that with this deformation the
ultraviolet divergence in the closed string emission rate is absent.
The linear dilaton CFT is simple enough, nontrivial yet, to be
treated exactly. There are many interesting related works. In
\cite{Kluson} the author discusses the D-brane decay in the linear
dilaton background from the viewpoint of the effective dynamics, see
also \cite{Nagami} with electromagnetic field turned on. In
\cite{Taka04} the author makes a deformation of 2d string theory
with the $X^0$ direction replaced by a timelike linear dilaton, and
discuss its matrix model description. The linear dilaton is also
related to many other important CFTs, e.g. the near throat region of
the NS5 brane, the 2d black hole $SL(2,\mathbb{R})/U(1)$ etc. The
brane decay in these backgrounds has been studied by \cite{NST,
Chen04, Rey04, Rey05} and other authors. Recently \cite{CSV05}
studies a background which is Minkowskian in the string frame with a
null linear dilaton, and proposes a dual matrix string model to
describe the dynamics near the big-bang singularity. A pure linear
dilaton has a bare strong coupling region. A convenient way to
regularize this in the string perturbation theory is to embed it
into the Liouville field theory, where the strong coupling region is
effectively screened by the exponential tachyon potential. In the
past few years we have gained much knowledge about the Liouville
theory, especially about the D-branes in this theory. There are two
types of D-branes. One is extended in the Liouville direction
\cite{FZZ, T}, the other is localized \cite{ZZ} in the strong
coupling region.

In this paper we study the decay process of a ZZ-type $p$-brane in
the background
\begin{eqnarray}
\mathbb{R}_t \times \mathbb{R}_L \times \mathbb{R}^{D-2}\,,
\end{eqnarray}
where the gradient of the dialton has components along
$\mathbb{R}_t$ and $\mathbb{R}_L$. The difference of our background
with that of \cite{UV} is that we turn on a exponential bulk tachyon
potential along $\mathbb{R}_L$, making it become a Liouville
direction. In presence of the linear dilaton it is illegal to impose
the usual Dirichlet boundary condition, since it breaks the
world-sheet conformal invariance. However having the exponential
potential we can utilize the knowledge from Liouville theory to
study the ZZ-type brane, which is localized in the strong coupling
region of $\mathbb{R}_L$ direction and has usual Dirichlet or
Neumann boundary conditions in other spacial directions. The time
direction CFT describing the brane decay is the so-called Timelike
Boundary Liouville (TBL) theory, which is initially studied in
\cite{TBL, ST03}. By tuning the gradient of the dilaton the
dimension $D$ of the spacetime varies from 2 to 26. The critical
case $D=26$ corresponds to the null dilaton. In this situation not
only the open string configuration on the D-brane is time-dependent,
the background in the bulk is also variant along with time.

This paper is organized as follows. In section 2 we introduce the
background and review the construction of the boundary states in
this background. In section 3 we calculate the closed string field
produced by the brane decay. The result is, as expected, that the
contribution of the linear dilaton changes the on-shell condition of
the closed string field. In section 4 we calculate the imaginary
part of the annulus amplitude, which is identified, using the
optical theorem, as the emission rate of closed strings. We find
that when $2<D<26$ (spacelike dilaton) there is a Hagedorn behavior
in the closed string UV, which is as same as the brane decay in the
usual 26d and 2d string theory \cite{Lambert, KMS03}, while
different from \cite{UV}, although there is also a linear dilaton.
Our result, as in \cite{Lambert}, means that all of the brane energy
converts into closed strings. In the closed string IR, however, the
emission rate is finite. These two aspects are the main results of
this paper. We also give some comments about the case $D=26$ which
corresponds that the dilaton is null. In section 5 we make some
concluding remarks. In appendix we give some calculation details.


\section{Background and boundary states}
In this paper we will study the decay process of a ZZ-type D-brane
in the framework of the boundary CFT. The total world-sheet action
is $S=S_{\mathrm{bulk}}+S_{\mathrm{bndy}}$. The bulk is a tensor
product of a time-like linear dilaton $X^0$, a Liouville direction
$X^1$ and the residual free bosons $X^i$ with $i=2,\cdots,D$. The
boundary action is the so-called ``half s-brane'' deformation of the
$X^0$ bulk CFT. To be explicit we list the total action
\begin{eqnarray}
S&=&\left(\,S_{X^0}+S_{X^1}+S_{X^i}\right)+S_{\mathrm{bndy}}\,,\nn\\[0.3cm]
S_{X^0}&=&-\frac{1}{4\pi}\int d^2\sigma\sqrt{g}\,(\,g^{ab}\p_aX^0\p_bX^0 + RV_0X^0)\,,\nn\\[0.2cm]
S_{X^1}&=& \frac{1}{4\pi}\int d^2\sigma\sqrt{g}\,(\,g^{ab}\p_aX^1\p_bX^1 + RV_1X^1 + 4\pi\mu e^{2bX^1})\,,\nn\\[0.2cm]
S_{X^i}&=& \frac{1}{4\pi}\int d^2\sigma\sqrt{g}\,g^{ab}\p_aX^i\p_bX^i\,,\nn\\[0.2cm]
S_{\mathrm{bndy}}&=&\frac{1}{2\pi}\int
ds\,g^{1/4}\,(\,KV_0X^0+2\pi\lambda\,e^{\beta X^0})\,.
\end{eqnarray}
Here $g_{ab}$ is the world-sheet metric, $R$ is the 2d curvature
scalar, and $K$ is the extrinsic curvature of the world-sheet
boundary. From the action we see that the dilaton
$\Phi=V_0X^0+V_1X^1$. The $X^0$ part is the Timelike Boundary
Liouville (TBL) with vanishing 2d cosmological constant,
introduced in \cite{TBL}. For the theory to be perturbatively
well-defined we also turn on the bulk tachyon $T\sim e^{2bX^1}$ to
make the $X^1$ direction to be a standard Liouville CFT. Conformal
invariance requires that
\begin{eqnarray}
V_0=\beta-\frac{1}{\beta}\,\,, \qquad V_1=b+\frac{1}{b}\,\,.
\end{eqnarray}
The central charge is
\begin{eqnarray}
c^{X^0}=1-6V_0^2 \,\,, \qquad c^{X^1}=1+6V_1^2 \,\,.
\end{eqnarray}
The dimension $D$ of spacetime is
\begin{eqnarray}
D=26-6(-V_0^2+V_1^2)\,.
\end{eqnarray}
The primary fields of the $X^0$ and $X^1$ CFTs are
$e^{ik_0X^0+V_0X^0}$ and $e^{ik_1X^1+V_1X^1}$, with the conformal
weights $-\frac{1}{4}(k_0^2+V_0^2)$ and $\frac{1}{4}(k_1^2+V_1^2)$
respectively.

To study the brane decay a central object is the boundary state of
the unstable brane. Since the total CFT is a tensor product of
several parts, we can construct the boundary state for each part
separately, and then multiply them together. We first deal with
the $X^0$ direction. The $X^0$ CFT is the so-called Timelike
Boundary Liouville (TBL) theory. The one point function and the
corresponding boundary state have been worked out in \cite{TBL}.
The TBL can be related to the spacelike boundary Lioville (SBL),
defined by
\begin{eqnarray}
\frac{1}{4\pi}\int d^2\sigma\sqrt{g}\,(\,g^{ab}\p_a\phi\,\p_b\phi
+ RQ\phi + 4\pi\mu e^{2b\phi}) +\frac{1}{2\pi}\int
ds\,g^{1/4}\,(\,KQ\phi+2\pi\lambda\,e^{b\phi})
\end{eqnarray}
with $Q=b+\frac{1}{b}$, by the analytical
continuation\footnote{Although the continuation from TBL to SBL
has some subtleties \cite{ST03,Sch1,Sch2}, especially in the
multi-point function, the naive manipulation in the one point
function gives the correct result.}
\begin{eqnarray}\label{Cont.}
X^0\rightarrow i\phi\,,\quad \beta\rightarrow-ib\,,\quad
V_0\rightarrow-iQ\,.
\end{eqnarray}
Then the FZZT one point function\footnote{Here we use the
unnormalized one point function, which is just the inner product
of the boundary state with the Ishibashi state.} obtained in
\cite{FZZ,T} gives us its timelike counterpart
\begin{eqnarray}
U_{X^0}(k_0)&=&\langle\,e^{ik_0X^0(0)+V_0X^0(0)}\rangle_{\mathrm{TBL}}=
   i\langle\,e^{-k_0\phi(0)+Q\phi(0)}\rangle_{\mathrm{SBL}} \nn\\[0.2cm]
&=&-\frac{i}{2^{1/4}2\pi
b}\left[\frac{2\pi\lambda}{\Gamma(1-b^2)}\right]^{\frac{k_0}{b}}
   \Gamma(-k_0/b)\,\Gamma(1-bk_0) \nn\\[0.1cm]
&=&-\frac{1}{2^{1/4}2\pi\beta}\left[\frac{2\pi\lambda}{\Gamma(1+\beta^2)}\right]^{-\frac{ik_0}{\beta}}
   \Gamma(ik_0/\beta)\,\Gamma(1-i\beta k_0) \,.
\end{eqnarray}
In the second line we have taken the limit (see \cite{TBL})
\begin{eqnarray}\label{mu=0}
\mu\rightarrow0\,, \quad s\rightarrow\infty\,,\quad \mu\cosh^2\pi
bs=\lambda^2\sin\pi b^2 \quad \mathrm{fixed}
\end{eqnarray}
to turn off the bulk cosmological constant term in the $X^0$
direction. Having the one point function we can construct the
corresponding boundary state as
\begin{eqnarray}\label{B0}
|B\r_{X^0}=\int\frac{dk_0}{2\pi}\,U_{X^0}(k_0)\,|k_0\r\r \,,
\end{eqnarray}
where $|k_0\r\r$ is the Ishibashi state corresponding to the
primary state $|k_0\r=e^{ik_0X^0+V_0X^0}|0\r$.

Now we turn to the $X^1$ part, which is the standard Liouville
theory. In this theory there is an important boundary state
\cite{ZZ}, called ZZ brane, corresponding to the degenerate
representation of the Virasoro algebra. The ZZ brane boundary
state is
\begin{eqnarray}\label{B1}
|B\r_{X^1}=\int\frac{dk_1}{2\pi}\,U_{X^1}(k_1)\,|k_1\r\r \,,
\end{eqnarray}
The unnormalized one point function is
\begin{eqnarray}
&&U_{X^1}(k_1)\,=\,\l\,e^{ik_1X^1(0)+V_1X^1(0)}\r_{ZZ}\,=\,
   -\frac{2^{3/4}\pi b\,\tilde{\mu}^{-ik_1/2b}}{\,\Gamma(-ik_1/b)\,\Gamma(1-ibk_1)}\,\,, \nn\\[0.2cm]
&&\tilde{\mu}\,=\,\pi\mu\gamma(b^2)\,,\quad
\gamma(b^2)\,=\,\Gamma(b^2)/\Gamma(1-b^2)\,,
\end{eqnarray}
where $|k_1\r\r$ is the Ishibashi state corresponding to the
primary state $|k_1\r=e^{ik_1X^1+V_1X^1}|0\r$.

The remaining parts of the boundary state corresponding to the
$X^i$ and the Fadeev-Popov ghost are standard
\begin{eqnarray}
|B\r_{X^i}&=&\int\frac{d^{D-p-2}k_\bot}{(2\pi)^{D-p-2}}\,
    \exp\left[-\sum_{n=1}^\infty\frac{1}{n}S_{ij}\,\alpha_{-n}^i\tilde{\alpha}_{-n}^j\right]
    |k_\bot,k_\|=0\r\,,\label{Bi}\\[0.2cm]
|B\r_{\mathrm{gh}}&=&\frac{1}{2}\,\exp\left[\sum_{n=1}^\infty(c_{-n}\tilde{b}_{-n}-b_{-n}\tilde{c}_{-n})\right]
    (c_0+\tilde{c}_0)\,|\downarrow\r\,.\label{Bgh}
\end{eqnarray}
The momentum integration in the $X^i$ part boundary state is only
over the transverse direction $k_\bot$. The ghost vacuum
$|\downarrow\r$ is defined by
$c_m|\downarrow\r=b_n|\downarrow\r=0$ for $m>0,n\geq0$.

The total boundary state $|B_p\r$ for the open string tachyon
condensation on a ZZ-type D$p$-brane is the tensor product of the
$X^0$, $X^1$, $X^i$ and the ghost part
\begin{eqnarray}\label{B}
|B_p\r=|B\r_{X^0}\otimes|B\r_{X^1}\otimes|B\r_{X^i}\otimes|B\r_{\mathrm{gh}}\,.
\end{eqnarray}

\section{Closed string field configuration} After reviewing the
construction of the boundary state in the previous subsection, we
now calculate the closed string field configuration produced by the
ZZ-type D$p$-brane, using the method of \cite{Sen04}. The closed
string field $|\Psi\r$ is a state with ghost number 2 in the Hilbert
space of the first quantized closed string theory, which satisfies
the constraints
\begin{eqnarray}
(b_0-\tilde{b}_0)|\Psi\r=0\,,\quad (L_0-\tilde{L}_0)|\Psi\r=0\,.
\end{eqnarray}
The linearized equation of motion of $|\Psi\r$ is
\begin{eqnarray}
2\,(\,Q_\mathrm{B}+\tilde{Q}_\mathrm{B})|\Psi\r=|B_p\r\,.
\end{eqnarray}
Here $Q_\mathrm{B}+\tilde{Q}_\mathrm{B}$ is the BRST charge. The
string field is easier to be found if we impose the Siegel gauge
condition
\begin{eqnarray}
(b_0+\tilde{b}_0)|\Psi\r=0\,.
\end{eqnarray}
Since the exactness of the Virasoro generator in the BRST
cohomology
\begin{eqnarray}
\{Q_\mathrm{B}+\tilde{Q}_\mathrm{B}\,,\,b_0+\tilde{b}_0\}=\,L_0+\tilde{L}_0\,,
\end{eqnarray}
we get the equation of motion in the Siegel gauge
\begin{eqnarray}\label{EOM(S)}
2\,(L_0+\tilde{L}_0)|\Psi\r=(b_0+\tilde{b}_0)|B_p\r\,.
\end{eqnarray}
Now using the expressions for the boundary state $|B_p\r$, we see
that the source term $(b_0+\tilde{b}_0)|B\r$ for the closed string
field, due to
$(b_0+\tilde{b}_0)(c_0+\tilde{c}_0)|\downarrow\r=2\,|\downarrow\r$,
is
\begin{eqnarray}\label{BB}
&&\int\frac{dk_0}{2\pi}\frac{dk_1}{2\pi}\frac{d^{D-p-2}k_\bot}{(2\pi)^{D-p-2}}\,
    U_{X^0}(k_0)|k_0\r\r\otimes U_{X^1}(k_1)|k_1\r\r   \nn\\
&&  \qquad
    \otimes\,\exp\left[-\sum_{n=1}^\infty\frac{1}{n}S_{ij}\,\alpha_{-n}^i\tilde{\alpha}_{-n}^j\right]|k_i\r\otimes
    \exp\left[\sum_{n=1}^\infty(c_{-n}\tilde{b}_{-n}-b_{-n}\tilde{c}_{-n})\right]|\downarrow\r\nn\\[0.2cm]
&\equiv&\int\frac{dk_0}{2\pi}\frac{dk_1}{2\pi}\frac{d^{D-p-2}k_\bot}{(2\pi)^{D-p-2}}\,
    U_{X^0}(k_0)\,U_{X^1}(k_1)\sum_{N=0}^\infty\hat{\mathcal{O}}_N\,|k_0,k_1,k_\bot,k_\|=0;\downarrow\r\,.
\end{eqnarray}
Here we group the various descendant states according to their
level, and introduce the operator $\hat{\mathcal{O}}_N$ to map the
primary state to a level $N$ state. This operator will in general
depend on the momentum $k_1$, due to the nontriviality of the
Liouville theory along the $X^1$ direction.

To solve the closed string field $|\Psi\r$ produced by the brane
decay, we make the following ansatz
\begin{eqnarray}
|\Psi\r=\int\frac{dk_0}{2\pi}\frac{dk_1}{2\pi}\frac{d^{D-p-2}k_\bot}{(2\pi)^{D-p-2}}\,
     U_{X^1}(k_1)\sum_{N=0}^\infty\hat{\mathcal{O}}_N\,\phi_N(k)\,|k,\downarrow\r\,,
\end{eqnarray}
where the symbol $k$ is a abbreviation of
$(k_0,k_1,k_\bot,k_\|=0)$, and $\phi_N(k)$ is the unknown function
to be determined. To calculate the action of $L_0$ and
$\tilde{L}_0$ on the closed string field $|\Psi\r$, we can using
the general argument in CFT to get the result, without knowing the
detail form of the descendant operator $\hat{\mathcal{O}}_N$
introduced above. Consider a Ishibashi state $|h\r\r$ associated
to the primary state $|h\r$, which can be written as
\begin{eqnarray}
|h\rangle\rangle=\sum_{N=0}^\infty\hat{\mathcal{O}}_N |h\rangle\,.
\end{eqnarray}
The index $N$ denotes the level, and the state
$\hat{\mathcal{O}}_N|h\rangle$ is a combination of level $N$
descendant states. It is not difficult to see that
\begin{eqnarray}
L_0\hat{\mathcal{O}}_N|h\rangle=\left(\sum_{i}h^i+N\right)\hat{\mathcal{O}}_N|h\rangle\,.
\end{eqnarray}
The action of $\tilde{L}_0$ is similar. Then we obtain the result of
$2(L_0+\tilde{L}_0)|\Psi\r$ as following
\begin{eqnarray}
\sum_{N=0}^\infty\int\frac{dk_0}{2\pi}\frac{dk_1}{2\pi}\frac{d^{D-p-2}k_\bot}{(2\pi)^{D-p-2}}\,\,
     \delta^p(k_\|)\,U_{X^1}(k_1)\left[\,k^2+V^2+4(N-1)\,\right]\phi_N(k)\,\hat{\mathcal{O}}_N|k,\downarrow\r\,.
\end{eqnarray}
Insert the above formula and (\ref{BB}) into the Siegel gauge EOM
(\ref{EOM(S)}) we have
\begin{eqnarray}
\left[\,k^2+V^2+4(N-1)\,\right]\phi_N(k)=\,U_{X^0}(k_0)\,.
\end{eqnarray}
Here we have dropped the factor $\delta^p(k_\|)$ and restricted to
$k_\|=0$. So we see that although the $X^1$ direction is a
nontrivial Liouville CFT, the result is simple: The equation we
need to determine the closed string field is just the same as the
usual free CFT rolling tachyon, except the contribution, the $V^2$
term, from the linear dilaton. Having the closed string field in
momentum space, we make the Fourier transformation with respect to
$k_0$ to get the $\phi_N(x^0,\mathbf{k})$ as
\begin{eqnarray}
\phi_N(x^0,\mathbf{k})&=&\int\frac{dk_0}{2\pi}\,\frac{U_{X^0}(k_0)}{\,-k_0^2+\omega_{\bf{k}}^2}\,e^{ik_0x^0}\nn\\[0.2cm]
    &=&\frac{\,i\,C}{\omega_{\bf{k}}}\,\tilde{\lambda}^{i\omega_{\bf{k}}/\beta}\,
    \Gamma(-i\omega_{\bf{k}}/\beta)\,\Gamma(1+i\beta\omega_{\bf{k}})\,e^{-i\omega_{\bf{k}}x^0}\,,\label{Fourier}
\end{eqnarray}
where $C$ is an unimportant constant,
$\omega_{\bf{k}}=\bf{k}^2+V^2+4(N-1)$,
$\tilde{\lambda}=2\pi\lambda/\Gamma(1+\beta^2)$. This is the
negative frequency solution. The positive one can be obtain by the
replacement $\omega\rightarrow-\omega$. This result can be
formally argued by use of the residue theorem. The positive and
negative frequency parts correspond to the different choice of the
contour. However the behavior of the integration along the large
semicircle is not easy to analyze. In Appendix we give a direct
calculation of this Fourier transformation.


\section{Closed string emission from the brane decay}
In this section we calculate the closed string emission following
the method of \cite{UV}. We first calculate the open string
partition function, which is just the product of each directions.
The optical theorem tells us that the imaginary part of this
partition is the closed string emission rate.

\subsection{Partition function}
The total open string partition function can be written as
\begin{eqnarray}
Z(t)=\l{B_p|\,\tilde{q}^{\,L_0+\tilde{L}_0-\frac{D-2}{12}}|B_p}\r\,,\quad\quad
\tilde{q}=e^{-4t}\,.
\end{eqnarray}
In this boundary state formalism, the annulus is viewed as the
propagation of a closed string, so the closed string Hamiltonian
$L_0+\tilde{L}_0-\frac{c}{12}$ appears in the exponential. Note
that we have included the ghost contribution which cancel two of
the total directions. This partition function, of course,
factorizes into the part of each direction. The nontrivial ones
are the $X^0$ and $X^1$ directions, the remaining parts, including
the ghost, are standard. The $X^0$ part of the partition function
is \cite{UV}
\begin{eqnarray}
Z_{X^0}(t)&=&\frac{\pi}{2}\int_{-\infty}^{\infty}dk_0\,\,
        \frac{\chi_{(k_0+iV_0)/2}(t)}{\,\sinh(\pi\beta{k_0})\sinh(\pi{k_0}/\beta)}\,\,,\\[0.2cm]
\chi_\alpha(t)&=&\eta(2it/\pi)^{-1}\,\tilde{q}^{\,-(\alpha-iV_0/2)^2}\,.
\end{eqnarray}
Here $\chi_\alpha$ is the character. After a modular
transformation to the open string channel and integrating over
$k_0$, the partition function becomes
\begin{eqnarray}\label{Z0}
Z_{X^0}(t)=\sqrt{2}\,\,\eta(i\pi/2t)^{-1}\,\int_{-\infty}^{\infty}d\nu\,\,q^{\nu^2}\,\frac{\,\p{f_\beta}}{\p\nu}(\nu)\,,
\quad\quad q=e^{-\pi^2/t}\,.
\end{eqnarray}
where the function $f_\beta(\nu)$ is defined as
\begin{eqnarray}
f_\beta(\nu)=\frac{1}{2}\int_{-\infty}^{\infty}\frac{dk_0}{k_0}
       \left[\frac{\sinh(\nu{k_0})}{\,\sinh(\beta{k_0})\sinh(k_0/\beta)}-\frac{\nu}{k_0}\right]\,.
\end{eqnarray}
Actually $f_\beta(\nu)$ here is just the the  special function
$\log{S_\beta}(\frac{\beta+1/\beta}{2}-\nu)$, which is widely used
in the literature of Liouville theory. The derivative of $f$ has
simple poles at
\begin{eqnarray}
\nu=\left(m+\frac{1}{2}\right)\beta+\left(n+\frac{1}{2}\right)/\,\beta\,,\qquad
m,n\in\mathbb{Z}\,.
\end{eqnarray}
The integration contour is chosen to go below the real axis for
negative $\nu$ and above for positive $\nu$. For the calculation
of the closed string emission rate it is needed to know the
imaginary part of $Z_{X^0}$, which arises when going around the
poles of $f_\beta(\nu)$
\begin{eqnarray}
\mathrm{Im}Z_{X^0}(t)=2\sqrt{2}\,\pi\,\,\eta(i\pi/2t)^{-1}\sum_{m,n=0}^{\infty}
   e^{-[(n+\frac{1}{2})\beta+(m+\frac{1}{2})/\beta]^2\pi^2/t}\,.
\end{eqnarray}
Now we turn to the partition function of the $X^1$ direction. The
ZZ brane \cite{ZZ} in this direction contains only the identity
operator and its descendant fields. It corresponds to a degenerate
representation of the Virasoro algebra. For general $b$ there is
only one null state at level one, so the character reads simply as
\begin{eqnarray}
\chi_{_{ZZ}}(q)=\eta(i\pi/2t)^{-1}\,\left[\,q^{-(b+1/b)^2/4}-q^{-(b-1/b)^2/4}\right]\,,\qquad
q=e^{-\pi^2/t}\,.
\end{eqnarray}
The partition function of the $X^1$ direction, in the open string
channel, is just the corresponding character
\begin{eqnarray}
Z_{X^1}(t)\,=\,\eta(i\pi/2t)^{-1}\,\left[\,q^{-(b+1/b)^2/4}-q^{-(b-1/b)^2/4}\right]\,,
\end{eqnarray}
since there is only one conformal family which corresponds to the
identity operator. This $Z_{X^1}$ is real, so it contributes to
the imaginary part of the total partition function just a
multiplicative factor. The remaining spacial part and the ghost
part of the partition function is standard
\begin{eqnarray}
Z_{X^i,\,\mathrm{gh}}(t)=\frac{V_p}{2\sqrt{2}\,\pi t}\int\frac{d^pk_\|}{(2\pi)^p}\,
   q^{k_\|^2}\,\eta(i\pi/2t)^{4-D}\,.
\end{eqnarray}
Therefore the total partition function, the product of each
part, reads as
\begin{eqnarray}
Z(t)&=&Z_{X^0}(t)\,Z_{X^1}(t)\,Z_{X^i,\,\mathrm{gh}}(t)\\[0.2cm]
    &=&\frac{V_p}{2\pi t}\,\eta(i\pi/2t)^{2-D}\left[\,q^{-(b+1/b)^2/4}-q^{-(b-1/b)^2/4}\right]
       \int_{-\infty}^{\infty}\hspace{-0.2cm}d\nu\int\hspace{-0.1cm}\frac{d^pk_\|}{(2\pi)^p}\,
       q^{\nu^2+k_\|^2}\frac{\,\p f_\beta}{\p\nu}(\nu)\,.\nn
\end{eqnarray}
The imaginary part of this partition function is
\begin{eqnarray}
\mathrm{Im}Z(t)&=&\frac{V_p}{t}\,\,\eta(i\pi/2t)^{2-D}\left[\,q^{-(b+1/b)^2/4}-q^{-(b-1/b)^2/4}\right]\nn\\
 &&\qquad\qquad
        \times\int\hspace{-0.1cm}\frac{d^pk_\|}{(2\pi)^p}\,q^{k_\|^2}\sum_{m,n=0}^{\infty}
        e^{-[(n+\frac{1}{2})\beta+(m+\frac{1}{2})/\beta]^2\pi^2/t}\,.
\end{eqnarray}

\subsection{Closed string emission}
In this subsection we will calculate the closed string emission rate
by use of the optical theorem. The annulus diagram can be cut open
along a circle, the unitarity tells us that $\mathrm{Im}\mathbf{Z}$,
the imaginary part of the annulus amplitude, is just the closed
string emission rate $\bar{N}$, which is what we want to know. The
imaginary part of the annulus amplitude can be easily obtained from
the CFT result in the previous subsection by integrating the moduli,
since the (perturbative) string theory is just the world-sheet CFT
coupled to the 2d gravity.
\begin{eqnarray}\label{ImZ}
\mathrm{Im}\mathbf{Z}
&=&V_p\int_0^\infty\frac{ds}{s}\,\frac{1}{(4\pi s)^{\,p/2}}\left[e^{s(b+1/b)^2/4}-e^{s(b-1/b)^2/4}\right]\,
   \eta(is/2\pi)^{2-D}\nn\\
& &\qquad\qquad\qquad\qquad\qquad\qquad
   \times\sum_{m,n=0}^{\infty}e^{-s[(n+\frac{1}{2})\beta+(m+\frac{1}{2})/\beta]^2}\,.
\end{eqnarray}
Here we have  integrated out the longitudinal momentum $k_\|$, and
made a coordinate transformation $t=\pi^2/s$. The variable $s$ is
the world-sheet time of the open string, while $t$ is that of the
closed string.

Now we analyze the potential divergence of the emission rate
$\mathrm{Im}\mathbf{Z}$. For the limit $s\rightarrow\infty$, which
is the open string IR and closed string UV, the integrand
becomes
\begin{eqnarray}\label{IR-open}
\frac{V_p}{s}\,\frac{1}{(4\pi s)^{\,p/2}}\,\,
e^{\frac{1}{4}(b+1/b)^2s}\,e^{-\frac{1}{4}(\beta+1/\beta)^2s}\,e^{\frac{D-2}{24}s}
\end{eqnarray}
The world-sheet Weyl invariance tells us
\begin{eqnarray}
26=D+6(-V_0^2+V_1^2)\,,\quad\quad
V_0=\beta-\frac{1}{\beta}\,,\quad V_1=b+\frac{1}{b}\,.
\end{eqnarray}
So we have
\begin{eqnarray}\label{zero}
\frac{1}{4}\left(b+\frac{1}{b}\right)^2-\,\frac{1}{4}\left(\beta+\frac{1}{\beta}\right)^2+\,\frac{D-2}{24}\,=\,\,0\,.
\end{eqnarray}
All of the exponential factors disappear. Therefore the
integration of (\ref{IR-open}) at the neighborhood of infinity is
convergent for all larger $p$ except $p=0$. Since this conclusion
directly follows from the world-sheet Weyl invariance, it does not
matter whether the dilaton is spacelike, timelike or lightlike.

There is another way to understand this facts from the picture of
the Euclidean D-brane. Notice that the expression (\ref{ImZ}) can
be viewed as the partition function of open strings stretched
between an array of Euclidean D-branes along imaginary time
\cite{Lambert, ImD}, i.e.
\begin{eqnarray}
\bar{N}\,=\,\mathrm{Im}\mathbf{Z}
       \,=\,V_p\,\left\langle B_{-}\left|\,\frac{b_0^+c_0^+}{L_0+\tilde{L}_0}\,\right|B_{+}\right\rangle\,,
\end{eqnarray}
where the boundary state $|B_+\rangle$ describes branes located at
imaginary time $X^0=i(n+\frac{1}{2})\beta$ for $n\geq0$, while
$|B_-\rangle$ denotes branes at $X^0=-i(m+\frac{1}{2})/\beta$ for
$m\geq0$. The other directions of these boundary state are the
tensor product of the ZZ brane with the free part. A closed string
UV divergence relates, through the modular transformation, with
the open string IR divergence. From this open string point of view
the equation (\ref{zero}) just says that the lightest state of the
open strings, stretched along imaginary time direction, is
massless.
\begin{eqnarray}
M_\mathrm{open}^2\,=\,\left(\frac{\beta}{2}+\frac{1}{2\beta}\right)^2
   -\,\frac{1}{4}\left(b+\frac{1}{b}\right)^2-\,\frac{D-2}{24}\,=\,0\,.
\end{eqnarray}
The first term is the energy due to the finite length between two
closest D-branes. The second term comes from the Liouville
direction. The last term is just the zero point energy.

To calculate the emitted energy we use the formula, presented in
\cite{Lambert},
\begin{eqnarray}
\bar{E}=\frac{\partial}{\partial a}\left\langle B\,\left|\,
        \left[\frac{b_0^+c_0^+}{L_0+\tilde{L}_0}\right]_\mathrm{ret}
        \right|B(a)\right\rangle_{a=0}\,.
\end{eqnarray}
It is not difficult to find that the emitted energy is finite for
$p>2$, while infinite for $p\leq2$. This result is the same as
that of \cite{Lambert}, while not the same as \cite{UV}. In
\cite{UV} although the linear dilaton is also tuned on, the moduli
integration in the closed string UV region is convergent
exponentially. This will raise a question. The original brane
tension is proportional to the inverse of the string coupling, so
the energy carried by the closed strings, which is finite, is
insufficient by a power of $g_s$ in the weak coupling limit. The
model studied here exhibits a same Hagedorn behavior as
\cite{Lambert}, due to the presence of the Liouville direction,
which contributes a exponentially increasing factor
$e^{s(b+1/b)^2/4}$. We focus on the particular case $p=0$, or the
$p$-brane with all extended spacial direction compactified on
circles. In this case the closed string emission rate diverges
logarithmically, and the emitted energy diverges linearly. It is
natural to chose the cutoff at $1/g_s$. Then we see that the
emitted energy has the same order of magnitude as the original
brane. The same conclusion as \cite{Lambert} follows: all of the
brane energy converts into outgoing closed strings, and most of
the energy is carried by closed strings of mass $\sim 1/g_s$.
These divergence seems to invalidate the classical open string
results. However we know that there is a new kind of open/closed
string duality \cite{Sen03} in the process of open closed string
tachyon condensation, according to which the complete dynamics of
an unstable D-brane is captured by the quantum open string theory
without any need to explicitly consider the coupling of the system
to closed strings. In the context of the two-dimensional string
theory, this duality can be checked more explicitly using the dual
matrix model. In the case studied here it is, however, difficult
to see it directly. However we believe that it is still right.

Next we go to the closed string IR region: $s\rightarrow0$. Consider
the following integral
\begin{eqnarray}\label{UV-open}
&&\sum_{m,n=0}^{\infty}\int_0^\delta\frac{ds}{s}\,\frac{1}{(4\pi s)^{\,p/2}}
      \left[e^{s(b+1/b)^2/4}-e^{s(b-1/b)^2/4}\right]\,e^{-sA_{mn}^2}\,\eta(is/2\pi)^{2-D}\nn\\
&\sim&\sum_{m,n=0}^{\infty}\int_0^\delta\frac{ds}{s}\,\frac{s}{(4\pi s)^{\,p/2}}\,\,
      e^{-sA_{mn}^2}\,\,\eta(is/2\pi)^{2-D}\,,\qquad s\rightarrow0\,,
\end{eqnarray}
where we have defined $A_{mn}\equiv(n+\frac{1}{2})\beta+(m+\frac{1}{2})/\beta$.
To analyze this expression it is convinient to take a modular transformation of the
Dedekind $\eta$-function to obtain the  following asymptotic behavior as $s\rightarrow0$:
\begin{eqnarray}
\eta(is/2\pi)^{2-D}\sim\left(\frac{s}{2\pi}\right)^{\frac{D-2}{2}}
\left(e^{\frac{(D-2)\pi^2}{6s}}+(D-2)\,e^{\frac{(D-26)\pi^2}{6s}}+\cdots\right)\,.
\end{eqnarray}
The first term corresponds to the closed string tachyon which is
an artifact of our bosonic string model and is absent in the
superstring theory, so, as usual, we simply ignore it. Insert the
second term into (\ref{UV-open}) we have
\begin{eqnarray}\label{A}
\sum_{m,n=0}^{\infty}\int_0^\delta\frac{ds}{(4\pi s)^{\,p/2}}\,\,
      \left(\frac{s}{2\pi}\right)^{\frac{D-2}{2}}(D-2)\,e^{\frac{(D-26)\pi^2}{6s}}\,e^{-sA_{mn}^2}\,.
\end{eqnarray}
When $D>2$, the above quantity is smaller than the following one with $\delta\rightarrow\infty$
\begin{eqnarray}
I=\sum_{m,n=0}^{\infty}\int_0^\infty\frac{ds}{(4\pi s)^{\,p/2}}\,\,
      \left(\frac{s}{2\pi}\right)^{\frac{D-2}{2}}(D-2)\,e^{\frac{(D-26)\pi^2}{6s}}\,e^{-sA_{mn}^2}\,,
\end{eqnarray}
since the integrand is positive. The infinite integral in $I$, after some trivial rescaling,
can be related to the modified Bessel function $K_\nu(z)$, which has the following integral representation
\begin{eqnarray}
K_\nu(z)=\frac{1}{2}\left(\frac{z}{2}\right)^\nu\int_0^\infty
         s^{-\nu-1}\exp\left(-s-\frac{z^2}{4s}\,\right)ds\,.
\end{eqnarray}
Then we have
\begin{eqnarray}
&&I=(D-2)\,C\sum_{m,n=0}^{\infty}A_{mn}^{^{-(D-p\,)/2}}K_{(D-p\,)/2}(z_{mn})\,,\\[0.2cm]
&&z_{mn}=\sqrt{\frac{2(26-D)}{3}}\,\pi A_{mn}\,,
\end{eqnarray}
where $C$ is an unimportant factor. To analyze the series $I$ to be convergent or not,
we need to know the behavior of the summand when $m,n$ is large. Notice that
when $m,n\rightarrow\infty$,
$z_{mn}$ also tends to infinity. Use the asymptotic expansion
$K_\nu(z)\sim Cz^{-1/2}\exp(-z)$ we know that the summand behaves as
\begin{eqnarray}
A_{mn}^{^{-(D-p-1\,)/2}}\exp\left[-\pi\sqrt{\frac{2(26-D)}{3}}\,A_{mn}\right]\,.
\end{eqnarray}
So the series $I$ is finite. Therefore we have proved that, for
$2<D<26$, there is no divergence in the emission rate when going to
the closed string IR region. This result is similar with \cite{UV},
since the spacial Liouville part of the partition function is not
important in the closed string IR region, while the time direction,
described by TBL, dominates here.


For the case of null dilaton, the dimension of the spacetime is
26. The IR behavior in the closed string channel is completely
different from that of the spacelike dilaton studied above. Set
$D=26$ in (\ref{A}), we need to estimate the following quantity
\begin{eqnarray}\label{A1}
24\,\sum_{m,n=0}^{\infty}\int_0^\delta\frac{ds}{(4\pi s)^{\,p/2}}\,\,
      \left(\frac{s}{2\pi}\right)^{12}\,e^{-sA_{mn}^2}\,.
\end{eqnarray}
The integral is essentially the incomplete Gamma function
\begin{eqnarray}
\gamma(\alpha,x)=\int_0^x e^t\,t^{\alpha-1}dt\,,
\end{eqnarray}
so (\ref{A1}) is equal to
\begin{eqnarray}
\mathrm{Const}.\times\sum_{m,n=0}^{\infty}A_{mn}^{\,\,-(26-p\,)/4}\,\,
\gamma\left(\frac{26-p}{2}\,,\,A_{mn}^{1/2}\delta\,\right)\,.
\end{eqnarray}
We first take the limit $\delta\rightarrow0$. By use of the
expansion $\gamma(\alpha,x)\sim\frac{1}{\alpha}\,x^\alpha$ as
$x\rightarrow0$, the quantity (\ref{A1}) tends to
\begin{eqnarray}
\mathrm{Const}.\times\sum_{m,n=0}^{\infty}1\,,
\end{eqnarray}
which is badly divergent. This result is reasonable, since when
$D=26$  we have no more the exponential factor
$e^{(D-26)\pi^2/6s}$ in (\ref{A}), which suppress the integrand
greatly as $s\rightarrow0$. Physically when the dilaton is null,
not only is the field configuration on the brane time-dependent,
the background in the bulk is also variant along with time. To
some extent the IR divergence we just find reflects this double
time dependence.


\section{Concluding remarks}
In presence of the linear dilaton a more natural treatment is to
embed it into the Liouville filed theory, especially when studying
the behavior of D-branes. It is impossible to impose the usual
Dirichlet boundary condition in the direction where the linear
dilaton tuned on, since it is incompatible with the world-sheet
conformal invariance. While when embedding into the Liouville
theory, we can talk about the extended FZZT brane and the localized
ZZ brane. The latter one is more interesting. The open string
dynamics on it gives a holographic description of the bulk physics
in two-dimensional string theory.

In this paper we consider the decay of the ZZ-type D$p$-brane in the
linear dilaton background with a Liouville potential switched on.
This kind of branes satisfy the ZZ boundary condition in the
Liouville direction and usual Dirichlet or Neumannn in other spacial
directions. We calculate the closed string field produced by the
brane decay, and also analyze the emission rate of closed strings
during this decay process. We find that when $2<D<26$ (spacelike
dilaton) there is a Hagedorn behavior in the closed string UV
region, as same as both in 26d and 2d string theory in this region.
In the case of $p=0$ (or the $p$-brane with all extended spacial
directions wrapped on circles), the energy of the original brane
completely converts into the outgoing closed strings. Due to the
presence of the Liouville direction our result is different from
that of \cite{UV}, although both have a linear dilaton background.
On the other hand, when going to the closed string IR, the emission
rate is finite. In this region the time direction CFT dominates. For
the case of null dilaton, the UV behavior does not change. In the IR
region the result is, however, divergent.

There are some future directions. It is interesting to study the
same question from the viewpoint of the effective dynamics. This
may give us more insight into it. We can also turn on some
electric or magnetic fluxes on the brane and see what happens. The
electromagnetic field on the brane induces the conserved charges.
The conservation law provides some constraints to the process,
making it more controllable. It is also possible to compactify
some spacial directions and to study the effects of the winding
closed strings emitted out from the unstable brane.

\section*{Acknowledgements}
We would like to thank Professor Bin Chen and Professor Miao Li
for reading the draft and valuable comments.

\appendix
\section{Fourier transformation of the closed string field}
In this appendix we give the direct calculation of the following
Fourier transformation
\begin{eqnarray}
I=\int_{-\infty}^{\infty}\frac{dk_0}{2\pi}\,\frac{\Gamma(ik_0/\beta)\,\Gamma(1-i{\beta}k_0)}
    {\,-k_0^2+\omega^2}\,\,\tilde{\lambda}^{-ik_0/\beta}\,e^{ik_0x^0}\,.
\end{eqnarray}
Using the Schwinger proper time
\begin{eqnarray}\label{Sch1}
-\frac{1}{\,k_0^2-\omega^2}=i\int_0^{\infty}dt\,e^{it(k_0^2-\omega^2)}\,,
\end{eqnarray}
and the integral representation of Gamma function, we have
\begin{eqnarray}
I&=&i\int_{-\infty}^{\infty}\frac{dk_0}{2\pi}\,\tilde{\lambda}^{-ik_0/\beta}\,e^{ik_0x^0}
    \int_0^{\infty}dt\,e^{it(k_0^2-\omega^2)}\int_0^{\infty}ds\,e^{-s}s^{-1+ik_0/\beta}
    \int_0^{\infty}ds'\,e^{-s'}s^{ik_0/\beta}\nn\\[0.2cm]
 &=&i{\int\hspace{-0.2cm}\int}\,ds\,ds'\,e^{-s-s'}s^{-1}\int_0^{\infty}dt\,e^{-i\omega^2t}
    \int_{-\infty}^{\infty}\frac{dk_0}{2\pi}\,\,e^{itk_0^2+ivk_0}\,,
\end{eqnarray}
where
$\,\,v=x^0-\frac{1}{\beta}\log\tilde{\lambda}+\frac{1}{\beta}\log{s}-\beta\log{s'}\,.$
The integration over $k_0$, by completing the square, is the
Fresnel integration, and can be work out. Then
\begin{eqnarray}
I&=&i\,C{\int\hspace{-0.2cm}\int}\,ds\,ds'\,e^{-s-s'}s^{-1}
     \int_0^{\infty}t^{-1/2}\exp\left(-i\omega^2t-\frac{iv^2}{4t}\,\right)dt\,.
\end{eqnarray}
We do not care about the numerical factor, and just write it as
$C$. Fortunately the integration over $t$, by some trivial
rescaling of $t$, is just the integral representation of Hankel
function with order minus one-half
\begin{eqnarray}
\int_0^{\infty}t^{-1/2}\exp\left(-i\omega^2t-\frac{iv^2}{4t}\,\right)dt\,=\,
i{\pi}e^{-i\pi/4}\sqrt{\frac{v}{2\omega}}H_{-\frac{1}{2}}(-\omega\,v)\,=\,
\frac{C}{\omega}\,\,e^{-i\omega v}\,.
\end{eqnarray}
Now the integration $I$ can be completely worked out as
\begin{eqnarray}
I\,=\,\frac{i\,C}{\omega}\,\,\tilde{\lambda}^{i\omega/\beta}\,\Gamma(-i\omega/\beta)\,
    \Gamma(1+i\beta\omega)\,e^{-i\omega x^0}\,.
\end{eqnarray}
Of course we can use another Schwinger proper time representation,
different from (\ref{Sch1}), as follows
\begin{eqnarray}
-\frac{1}{\,k_0^2-\omega^2}=-i\int_0^{\infty}dt\,e^{-it(k_0^2-\omega^2)}\,,
\end{eqnarray}
The corresponding result is similar, with the replacement $\omega\rightarrow-\omega$,
\begin{eqnarray}
I'\,=\,\frac{i\,C}{\omega}\,\,\tilde{\lambda}^{-i\omega/\beta}\,\Gamma(\,i\omega/\beta)\,
    \Gamma(1-i\beta\omega)\,e^{i\omega x^0}\,.
\end{eqnarray}
These two result, $I$ and $I'$, correspond to the different choices of the boundary conditions.
One is the negative frequency solution, the other is the positive one.


\begin{thebibliography}{99}

\bibitem{Sen1}A. Sen, ``Rolling Tachyon,'' JHEP 0204 (2002) 048 [hep-th/0203211].
\bibitem{Sen2}A. Sen, ``Tachyon Matter,'' JHEP 0207 (2002) 065 [hep-th/0203265].
\bibitem{Sen3}A. Sen, ``Field Theory of Tachyon Matter,'' Mod. Phys. Lett. A17 (2002) 1797 [hep-th/0204143].
\bibitem{Review}A. Sen, ``Tachyon Dynamics in Open String Theory,'' Int. J. Mod. Phys. A20 (2005) 5513
              [hep-th/0410103].

\bibitem{Chen}B. Chen, M. Li and F.-L. Lin, ``Gravitational Radiation of Rolling Tachyon,'' JHEP 0211 (2002) 050
              [hep-th/0209222].
\bibitem{Larsen}F. Larsen, A. Naqvi and S. Terashima, ``Rolling Tachyons and Decaying Branes,''
              JHEP 0302 (2003) 039 [hep-th/0212248].
\bibitem{Rey}S.-J. Rey and S. Sugimoto, ``Rolling Tachyon with Electric and Magnetic Fields - T-duality approach,''
              Phys.Rev. D67 (2003) 086008 [hep-th/0301049].
\bibitem{Lambert}N. Lambert, H. Liu and J. Maldacena, ``Closed strings from decaying D-branes,'' [hep-th/0303139].
\bibitem{UV}J. L. Karczmarek, H. Liu, J. Maldacena and A. Strominger, ``UV Finite Brane Decay,''
            JHEP 0311 (2003) 042 [hep-th/0306132].
\bibitem{Sen03}A. Sen, ``Open-Closed Duality at Tree Level,''  Phys. Rev. Lett. 91 (2003) 181601 [hep-th/0306137].
\bibitem{Rey06}Y. Nakayama, S.-J. Rey and Y. Sugawara, ``Unitarity Meets Channel-Duality for Rolling /
               Decaying D-Branes,'' [hep-th/0605013].
\bibitem{McG03}J. McGreevy and H. Verlinde, ``Strings from Tachyons,'' JHEP 0312 (2003) 054 [hep-th/0304224].
\bibitem{KMS03}I. R. Klebanov, J. Maldacena and N. Seiberg, ``D-brane Decay in Two-Dimensional String Theory,''
               JHEP 0307 (2003) 045 [hep-th/0305159].
\bibitem{Taka03}T. Takayanagi and N. Toumbas, ``A Matrix Model Dual of Type 0B String Theory in Two Dimensions,''
               JHEP 0307 (2003) 064 [hep-th/0307083].
\bibitem{new-hat}M. R. Douglas, I. R. Klebanov, D. Kutasov, J. Maldacena, E. Martinec and N. Seiberg,
               ``A New Hat For The c=1 Matrix Model,'' [hep-th/0307195].
\bibitem{McMV}J. McGreevy, S. Murthy and H. Verlinde, ``Two-dimensional superstrings and the supersymmetric
              matrix model,'' JHEP 0404 (2004) 015 [hep-th/0308105].
\bibitem{Kluson}J. Kluson, ``Note on D-Brane Effective Action in the Linear Dilaton Background,''
              JHEP 0311 (2003) 068 [hep-th/0310066]. ``Bosonic D-brane Effective Action in Linear
              Dilaton Background,'' JHEP 0402 (2004) 024 [hep-th/0401236]. ``Proposal for the Open String
              Tachyon Effective Action in the Linear Dilaton Background,'' JHEP 0406 (2004) 021 [hep-th/0403124].
\bibitem{Nagami}K. Nagami, ``Rolling Tachyon with Electromagnetic Field in Linear Dilaton Background,''
              Phys. Lett. B591 (2004) 187 [hep-th/0312149].
\bibitem{Taka04}T. Takayanagi, ``Matrix Model and Time-like Linear Dilaton Matter,''
              JHEP 0412 (2004) 071 [hep-th/0411019].
\bibitem{NST}Y. Nakayama, Y. Sugawara and H. Takayanagi, ``Boundary States for the Rolling D-branes in NS5 Background,''
             JHEP 0407 (2004) 020 [hep-th/0406173].
\bibitem{Chen04}B. Chen, M. Li and B. Sun, ''Dbrane Near NS5-branes: with Electromagnetic Field,''
              JHEP 0412 (2004) 057 [hep-th/0412022].
\bibitem{Rey04}Y. Nakayama, K. L. Panigrahi, S.-J. Rey and H. Takayanagi, ``Rolling Down the Throat in
              NS5-brane Background: The Case of Electrified D-Brane,'' JHEP 0501 (2005) 052 [hep-th/0412038].
\bibitem{Rey05}Y. Nakayama, S.-J. Rey and Y. Sugawara, ``D-Brane Propagation in Two-Dimensional Black Hole Geometries,''
              JHEP 0509 (2005) 020 [hep-th/0507040].


\bibitem{CSV05}B. Craps, S. Sethi and E. Verlinde, ``A Matrix Big Bang,'' JHEP 0510 (2005) 005
              [hep-th/0506180].

\bibitem{FZZ}V. Fateev, A. Zamolodchikov and Al. Zamolodchikov, ``Boundary Liouville Field Theory I.
             Boundary State and Boundary Two-point Function,'' [hep-th/0001012].
\bibitem{T}J. Teschner, `` Remarks on Liouville Theory with Boundary,'' [hep-th/0009138].
\bibitem{ZZ}A. Zamolodchikov and Al. Zamolodchikov, ``Liouville Field Theory on a Pseudosphere,'' [hep-th/0101152].


\bibitem{TBL}M. Gutperle and A. Strominger, ``Timelike Boundary Liouville Theory,''
             Phys. Rev. D67 (2003) 126002 [hep-th/0301038].
\bibitem{ST03}A. Strominger and T. Takayanagi, ``Correlators in Timelike Bulk Liouville Theory,''
             Adv. Theor. Math. Phys. 7 (2003) 369 [hep-th/0303221].
\bibitem{Sch1}V. Schomerus, ``Rolling Tachyons from Liouville Theory,''  JHEP 0311 (2003) 043 [hep-th/0306026].
\bibitem{Sch2}S. Fredenhagen and V. Schomerus, ``Boundary Liouville Theory at c=1,''
             JHEP 0505 (2005) 025 [hep-th/0409256].
\bibitem{Sen04}A. Sen, ``Rolling Tachyon Boundary State, Conserved Charges and Two Dimensional String Theory,''
              JHEP 0405 (2004) 076 [hep-th/0402157].
\bibitem{ImD}D. Gaiotto, N. Itzhaki and L. Rastelli, ``Closed Strings as Imaginary D-branes,''
             Nucl. Phys. B688 (2004) 70 [hep-th/0304192].


\end{thebibliography}
\end{document}